\documentstyle[12pt]{article}

\catcode`\@=11
\def\marginnote#1{}
\newcount\hour
\newcount\minute
\newtoks\amorpm
\hour=\time\divide\hour by60
\minute=\time{\multiply\hour by60 \global\advance\minute
by-\hour}\edef\standardtime{{\ifnum\hour<12
\global\amorpm={am}%
        \else\global\amorpm={pm}\advance\hour by-12 \fi
        \ifnum\hour=0 \hour=12 \fi
        \number\hour:\ifnum\minute<10
0\fi\number\minute\the\amorpm}}
\edef\militarytime{\number\hour:\ifnum\minute<10
0\fi\number\minute}

\def\draftlabel#1{{\@bsphack\if@filesw {\let\thepage\relax
   \xdef\@gtempa{\write\@auxout{\string
      \newlabel{#1}{{\@currentlabel}{\thepage}}}}}\@gtempa
   \if@nobreak \ifvmode\nobreak\fi\fi\fi\@esphack}
        \gdef\@eqnlabel{#1}}
\def\@eqnlabel{}
\def\@vacuum{}
\def\draftmarginnote#1{\marginpar{\raggedright\scriptsize\tt#1}}
\def\draft{\oddsidemargin -.5truein
        \def\@oddfoot{\sl preliminary draft \hfil
        \rm\thepage\hfil\sl\today\quad\militarytime}
        \let\@evenfoot\@oddfoot \overfullrule 3pt
        \let\label=\draftlabel
        \let\marginnote=\draftmarginnote

\def\@eqnnum{(\theequation)\rlap{\kern\marginparsep\tt\@eqnlabel}%
\global\let\@eqnlabel\@vacuum}  }

\def\numberbysection{\@addtoreset{equation}{section}
        \def\theequation{\thesection.\arabic{equation}}}

\def\underline#1{\relax\ifmmode\@@underline#1\else
 $\@@underline{\hbox{#1}}$\relax\fi}

\catcode`@=12
\relax

\def\fin{\end{document}}
\def\beq{\begin{equation}}
\def\eeq{\end{equation}}
\def\bea{\begin{eqnarray}}
\def\eea{\end{eqnarray}}

\def\Gcc#1 {{\cal G}^{[ #1 ]}_\|}
\def\Gcp#1 {{\cal G}^{[ #1 ]}_\bot}
\def\Cco#1 {{\cal C}^{[ #1 ]} }

\begin{document}

\begin{titlepage}
\nopagebreak
\begin{flushright}
LPTENS-94/23\\
SWAT/93-94/38\\
hep--th/9408042
\\
July 1994
\end{flushright}

\vglue 2.5  true cm
\begin{center}
{\large\bf
LOTKA--VOLTERRA TYPE EQUATIONS \\
AND THEIR EXPLICIT INTEGRATION}

\vglue 1  true cm
{\bf Jean-Loup~GERVAIS}\\
{\footnotesize Laboratoire de Physique Th\'eorique de
l'\'Ecole Normale Sup\'erieure\footnote{Unit\'e Propre du
Centre National de la Recherche Scientifique,
associ\'ee \`a l'\'Ecole Normale Sup\'erieure et \`a l'Universit\'e
de Paris-Sud.},\\
24 rue Lhomond, 75231 Paris C\'EDEX 05, ~France.}\\
{\bf and}\\
{\bf Mikhail V.  SAVELIEV\footnote{Permanent address:
 Institute for High Energy Physics,
142284, Protvino, Moscow region, Russia.}}\\
{\footnotesize Department of Mathematics,
University College of Swansea, \\ Swansea SA2 8PP, Wales, U.K.
 }\\

\medskip
\end{center}

\vfill
\begin{abstract}
\baselineskip .4 true cm
\noindent
In the present note we give an explicit integration
of some two--dimensionalised Lotka--Volterra
type equations associated with simple Lie algebras
possessing a representation without branching.

\end{abstract}
\vfill
\end{titlepage}
\baselineskip .5 true cm

Equations  of the Lotka--Volterra type under these or that boundary conditions
appear in various branches of theoretical and mathematical physics, in
particular in some solvable models of the field theory, plasma physics,
electrotechnique, ecology, mathematical biology, etc. In the present note
we discuss a special class of these equations associated with simple
finite--dimensional Lie algebras.
Namely, in paper \cite{S87}  the finite system of the
Lotka--Volterra type equations
\begin{eqnarray}
\partial _+ R_{2\alpha -1} & = & \lambda _{\alpha } e^{R_{2\alpha }}
- \lambda _{\alpha -1} e^{R_{2\alpha -2}},\nonumber\\
\partial _- R_{2\alpha } & = & - e^{R_{2\alpha +1}} +
e^{R_{2\alpha -1}},\; 1\leq \alpha \leq N-1;
\label{1}
\end{eqnarray}
was obtained for the functions $R_A(z_+, z_-),\, 1\leq A\leq 2N$;
with the boundary conditions $R_0=R_{2N+1}=-\infty$.
Here $\lambda _{\alpha }$ are some constants which can be eliminated, e.g. by
redefinition of the functions $R_{2\alpha }$, only for the case related
to the algebra $A_r$ when none of them equal zero, see below.
These equations realise, in fact, a relation between the Toda system
\begin{equation}
\partial _+\partial _-x_i=\exp \rho _i;\quad  \rho _i=\sum _{j=1}^r
k_{ij}x_j, \; 1\leq i\leq r;\label{2}
\end{equation}
associated with a complex simple Lie algebra ${\cal G}$ (of rank
$r$ with the Cartan matrix $k$) possessing a representation
of dimension $N$ without branching, and the $A_{N-2}\oplus {\cal
Gl}(1)$--Toda fields. Seemingly, these equations can be extended to the system
of $1$st order partial differential equations associated with the relevant
affine Lie algebras; that is rather important for constructing their soliton
solutions.

Recall that speaking about a representation
without branching, we mean such a representation of weight $l$, for which
the representation space, endowed e.g. with the Verma base constructed
with the help of the (cyclic) highest weight vector $v_l$ by the
action on it of the negative simple root vectors, $X_{-j_a}\cdots
X_{-j_1}v_l,\;
1\leq j\leq r,\, 1\leq a\leq N-1$, contains the basis vectors $v_l(j_1,
\cdots ,j_a)$ with the only one value $j_m$ corresponding to a
given integer $m$. It is known that the $1$st fundamental representations
of $A_r, B_r, C_r$ and $G_2$ possess this property.

Note that for the case of $A_r$ system (\ref{1}) realises the proper
B\"acklund transformation, and, correspondingly, coincides with the
two--dimensionalised difference KdV equations derived and completely
integrated in \cite{LSS80} in terms of the general solutions to the
$A_r$-Toda system, while in some different form than those of
our formula (\ref{4}) when ${\cal G}=A_r$.

Equations (\ref{1}) represent an equivalent form of writing the system
\begin{eqnarray}
\partial _-(e^{\theta _a}\Psi _a) & = & e^{\theta _a}\Psi _{a+1};\\
\label{3}
\partial _+ \Psi _a & = & \lambda _a \exp \rho _{j_a} \Psi _{a-1};
\label{4}
\end{eqnarray}
$\quad 0\leq a \leq N-1$, with the identification
\[
R_{2\alpha -1} \equiv \mbox{ log } (\Psi_{\alpha }/\Psi_{\alpha -1}),\;
R_{2\alpha } \equiv \rho _{j_{\alpha }} - \mbox{ log } (\Psi_{\alpha }/
\Psi_{\alpha -1}).\]
Here we make use the following notations:
\[
\lambda _a \equiv l_{j_a}\,- \sum _{q=1}^{a-1} k_{j_qj_a};\;
\theta _a =\sum _j(l_j\,- \sum _{q=1}^{a} k_{j_qj})x_j,\;
\rho _{j_a}=\sum _jk_{j_aj}x_j.\]

When ${\cal G}=A_r$, equations (\ref{3}) and (\ref{4}), in terms of the
functions $e^{\theta _{a-1}/2}\Psi_{a-1}$, are equivalent to the Frenet--Serret
formulas for the moving frame \cite{GM93}. For other simple Lie algebras we
expect the similar situation with respect to the $W$--surfaces introduced
in \cite{GS93}.

Due to their nilpotent structure, equations (\ref{3}) and (\ref{4})
can be solved in
general form explicitly. Let us start from the first set of
these equations, i.e. (\ref{3}). (Recall that $\Psi _N \equiv 0$ by
definition.)
Taking $a=0$, the corresponding equation is rewritten as
\[
\partial _- (e^{\theta _0}\,\Psi _0)=e^{\theta _0}\,\Psi _1,\]
thereof
\[
\Psi _1 \,= e^{-\theta _0}\partial _-(e^{\theta _0}\Psi _0),\]
where, in accordance with equation (\ref{4}) for $a=0$, the function $\Psi _0$
does not depend on $z_+$, $\Psi _0=\Psi _0(z_-)$.
Substituting this solution for $\Psi _1$ in the equation (\ref{3})
for $a=1$, i.e., in
\[
\partial _-(e^{\theta _1}\,\Psi _1)=
e^{\theta _1}\,\Psi _2,\]
we get
\[\Psi _2 \,= e^{-\theta _1}\partial _-(e^{\theta _1}\Psi _1)=
e^{-\theta _1}\partial _-(e^{\theta _1-\theta _0}\partial _-
(e^{\theta _0} \Psi _0)).\]
Continuing this integration procedure, we obtain the expressions
for the whole set of the functions $\Psi _a$, $a>0$, in terms of the
function $\Psi _0$. The last step in the integration scheme, corresponding
equation (\ref{3}) with $a=N-1$, gives the equation of the $N$-th order for the
function $\Psi _0$, the generalised Bargmann--Leznov
generating equation \cite{L84}.
This equation establish a relation between its coefficients
(which are nothing but the $W$--potentials or conserved
currents\footnote{ Note that a formulation of the elements of the corresponding
$W$--algebra in terms of the Toda fields satisfying
equations (\ref{2}) comes back to the paper \cite{BG89}, see also
\cite{O'R90}. At the same time, such differential polynomials in Toda fields
arise
also as the characteristic integrals providing the integrability property of
these
equations, see e.g. \cite{LS92}.}) and the set of linearly independent
fundamental
solutions $\Psi ^A_0(z_-)$, which are just the functions which
determine, together with the analogous set of the functions
$\Phi ^A_0(z_+)$, the
general solutions to the Toda system. The final result for the solutions $R_A$
of
system (\ref{1}) can be written as:
\begin{equation}
e^{R_{2\alpha -1}}=\partial _-\mbox{ log }
(e^{-\rho _{j_{\alpha -1}}}\partial _-
(\cdots \,e^{-\rho _{j_1}}
\partial _- \, (e^{\theta _0} \Psi _0))),
\label{5}
\end{equation}
and $R_{2\alpha}=\rho _{j_{\alpha}}-R_{2\alpha -1}$.
The corresponding expression for the functions $\Psi _a$ contains $a$
derivatives
over $z_-$, and represents a generalisation of the determinant
form of writing the formula for the moving frame for $A_r$ given in
\cite{GM93}. These solutions, being substituted in
equations (\ref{4}), automatically satisfy them with account of
the Toda system (\ref{2}).
The generating equation in the case under consideration here, can be written
in the form
\begin{equation}
(\partial _z +\partial _z\theta _{N-1})\cdots  (\partial _z +\partial _z\theta
_{1})\cdot
(\partial _z +\partial _z\theta _{0}) \cdot \Psi _0(z)=0.\label{6}
\end{equation}
This equation implies, in particular, the following
expressions for the elements of the corresponding
$W$--algebra:
\begin{equation}
W_p=\sum _{q_1=p}^{N}\sum _{q_2=p-1}^{q_1-1}\cdots \sum
_{q_p=1}^{q_{p-1}-1}{\cal D}^0\cdot {\cal D}^1
\cdots {\cal D}^{p-1}\cdot e^{\theta _{q_p-1}},\label{7}
\end{equation}
where ${\cal D}^p=\exp (\theta _{q_p-1}-\theta _{q_{p+1}-1})\cdot \partial _z$.

The authors are grateful to  L. N. Lipatov, D. I. Olive and A. V.
Razumov for useful discussions. This work was
partially supported by the Russian Fund for Fundamental
Research, Grant \# 94-01-01585-A; and ISF, Grant \# RMO 000.

\end{document}